\documentclass[twocolumn,showpacs,preprintnumbers,amsmath,amssymb,nofootinbib,longbibliography]{revtex4-1}

\usepackage{graphicx}
\usepackage{dcolumn}
\usepackage{bm}
\usepackage{color}
\usepackage{epstopdf}
\usepackage[colorlinks, citecolor=blue]{hyperref}
\usepackage{lipsum}
\usepackage[export]{adjustbox}

\newcommand{\fig}[1]{Fig.~\ref{#1}}

\begin{document}

\title{Dark states in spin-polarized transport through triple quantum dot molecules}

\author{K. Wrze\'sniewski}
\email{wrzesniewski@amu.edu.pl}

\author{I. Weymann}
\affiliation{Faculty of Physics, Adam Mickiewicz University, Umultowska 85, 61-614 Pozna{\'n}, Poland}


\begin{abstract}
We study the spin-polarized transport through a triple quantum dot molecule weakly coupled to ferromagnetic leads.
The analysis is performed by means of the real-time diagrammatic technique
including up to the second order of perturbation expansion with respect to the tunnel coupling.
The emphasis is put on the impact of dark states on spin-resolved transport characteristics.
It is shown that the interplay of coherent population trapping
and cotunneling processes results in a highly nontrivial behavior
of the tunnel magnetoresistance, which can take negative values.
Moreover, a super-Poissonian shot noise is found
in transport regimes where the current is blocked
by the formation of dark states, which can be additionally
enhanced by spin-dependence of tunneling processes,
depending on magnetic configuration of the device.
The mechanisms leading to those effects are thoroughly discussed.
\end{abstract}


\maketitle


\section{Introduction}

Vast progress in theoretical and experimental studies of artificial molecules,
such as those realized in coupled quantum dot systems,
ceaselessly brings about many attractive and relevant results and observations
\cite{kouwenhoven01,hanson07}.
On the one side, such nanostructures have promising applications
for quantum computation \cite{jozsa95, hawrylak97,loss98},
where spin-polarized electron encodes a qubit.
On the other side, coupled quantum dots
exhibit various promising transport phenomena \cite{kouvenhoven97, andergassen2010},
which may be important for novel spintronic and nanoelectronic devices.
A particularly prominent example of an artificial molecular nanostructure
is a system built of three quantum dots (TQD).

The properties of triple quantum dot systems have been extensively studied in various regimes and configurations,
exposing rich Kondo physics \cite{Avishai2003, Zitko2006, Numata2009, Lopez2013},
various transport effects and complex electron structure
\cite{korkusinski2007, Amaha2013, Sanchez2014,  Cheng2017, Glodzik2017, Zhang2017},
as well as revealing potential for applications in quantum computing
\cite{Laird2010, Gaudreau12, Luczak2014, Luczak2017, Russ2017}
and for generation of non-local, entangled electron pairs \cite{Fulop2015, dominguez2016}.
When the three quantum dots form a triangular geometry
\cite{Gaudreau06, Amaha2008, Mitchell2009, Bulka2011, hsieh12, seo2013, wrzesniewskiPRB15},
the system resembles a simple planar molecule and, due to the interference effects,
the formation of dark states is possible \cite{emaryPRB07, emaryPRB80, kostyrko09, weymannPRB83, Kubo2016, grifoni17}.
This quantum-mechanical phenomenon was first observed in atomic physics
\cite{Whitley1976, Boller1991, Fleischhauer2005, Xia2015},
and then found also in mesoscopic systems, such as, in particular, coupled quantum dots \cite{Beenakker2006, Bayer2008}.
A dark state emerges when destructive interference of electronic wavefunctions
decouples the system from one of the leads.
When the system is in a dark state,
it results in a coherent electron trapping  \cite{Beenakker2006}
and, consequently, a strong current suppression,
negative differential conductance and enhanced shot noise \cite{emaryPRB07, emaryPRB80, kostyrko09, weymannPRB83}.
Interestingly, the dark states in quantum dot systems
are also considered to enable the creation of spatially separated spin-entangled
two-electron states \cite{emaryPRB13}
and, thus, open the possibility to build various quantum logic devices
as well as quantum memory \cite{aharon2016}.
It is important to note that the mechanism of coherent population trapping
is very distinct from Coulomb, spin \cite{Ono2002, Fransson2006, Platero2013, Tarucha2017}
and Franck-Condon \cite{Leturcq2009} blockades
or Aharonov-Bohm \cite{Kuzmenko2006} effect on triangular quantum dots, to name a few.

All this provides a strong motivation for further considerations
of dark states in transport through triple quantum dots
and, in fact, there are still certain aspects that remain unexplored.
One of them involves the role of dark states
in spin-resolved transport behavior.
In fact, spin-dependent phenomena in transport through quantum dot systems
are currently intensively studied \cite{Awschalom2013Mar}.
This is not only due to expected applications for spintronics and spin nanoelectronics \cite{Zutic2004Apr},
but also because of the possibility to controllably
explore the fundamental interactions between single charges and spins \cite{Hanson2007Oct}.
First of all, the presence of ferromagnetic (FM) electrodes
introduces many qualitative and quantitative changes in transport,
resulting in magnetoresistive or spin diode-like behavior \cite{barnas98,Weymann05,barnas08}.
Moreover, the effects, such as the suppression of the Kondo effect with
the emergence of an exchange field \cite{Martinek03}, universal magneto-conductance scaling \cite{Gaass11},
an enhancement of splitting efficiency of entangled Cooper pairs in QD based splitters \cite{weymann14, wrzesniewski17}
or spin thermoelectric effects \cite{Swirkowicz2009Nov,Weymann2013Aug,Karwacki16}
are all among many interesting phenomena arising from the
coupling of quantum dots to FM leads.
In this context, however, the interplay between the coherent population trapping
and spin-resolved tunneling has so far hardly been explored.
Therefore, in this paper we address this problem and analyze the spin-dependent transport
through triple quantum dots weakly attached to two ferromagnetic electrodes,
focusing on the parameter regime where the dark states form.

To determine the nonequilibrium transport characteristics,
we use the real-time diagrammatic technique \cite{diagrams1},
including the first and second-order diagrams with respect to the tunnel coupling.
This allows us to systematically include the sequential and cotunneling
processes in the transport analysis.
We study the bias and gate voltage dependence of the current, differential conductance
and Fano factor in two different magnetic configurations
of the device: the parallel and antiparallel one.
Furthermore, by calculating the currents in the two magnetic configurations
we also determine the tunnel magnetoresistance (TMR) of the system
\cite{julliere, barnas98, Weymann05, barnas08}.
These quantities provide a relevant insight into the spin-dependent
transport properties of the considered system,
and are especially interesting in the regimes where dark states are present.
In particular, we focus on the transport regimes where
one and two particle dark states and their hole counterparts form.
We show that when the system is trapped in a dark state,
the current flows mainly due to cotunneling processes.
Moreover, depending on a particular type of dark state,
we find a strong dependence of the current
on magnetic configuration of the device,
which results in a nontrivial behavior of the TMR.
A similarly strong dependence on magnetic configuration
is also found in the case of shot noise, which
is generally super-Poissonian in the dark state regions.

The paper is structured as follows.
Section \ref{theoretical framework} consists of
model description and method used for numerical calculations.
In Sec. \ref{results} we present the numerical results and relevant analysis.
This section is divided into four subsections relating
to different types of examined dark states.
Finally, the work is concluded in Sec. \ref{conclusions}.


\section{Theoretical framework} \label{theoretical framework}

\subsection{Model}

\begin{figure}[t]
  \includegraphics[width=0.95\columnwidth]{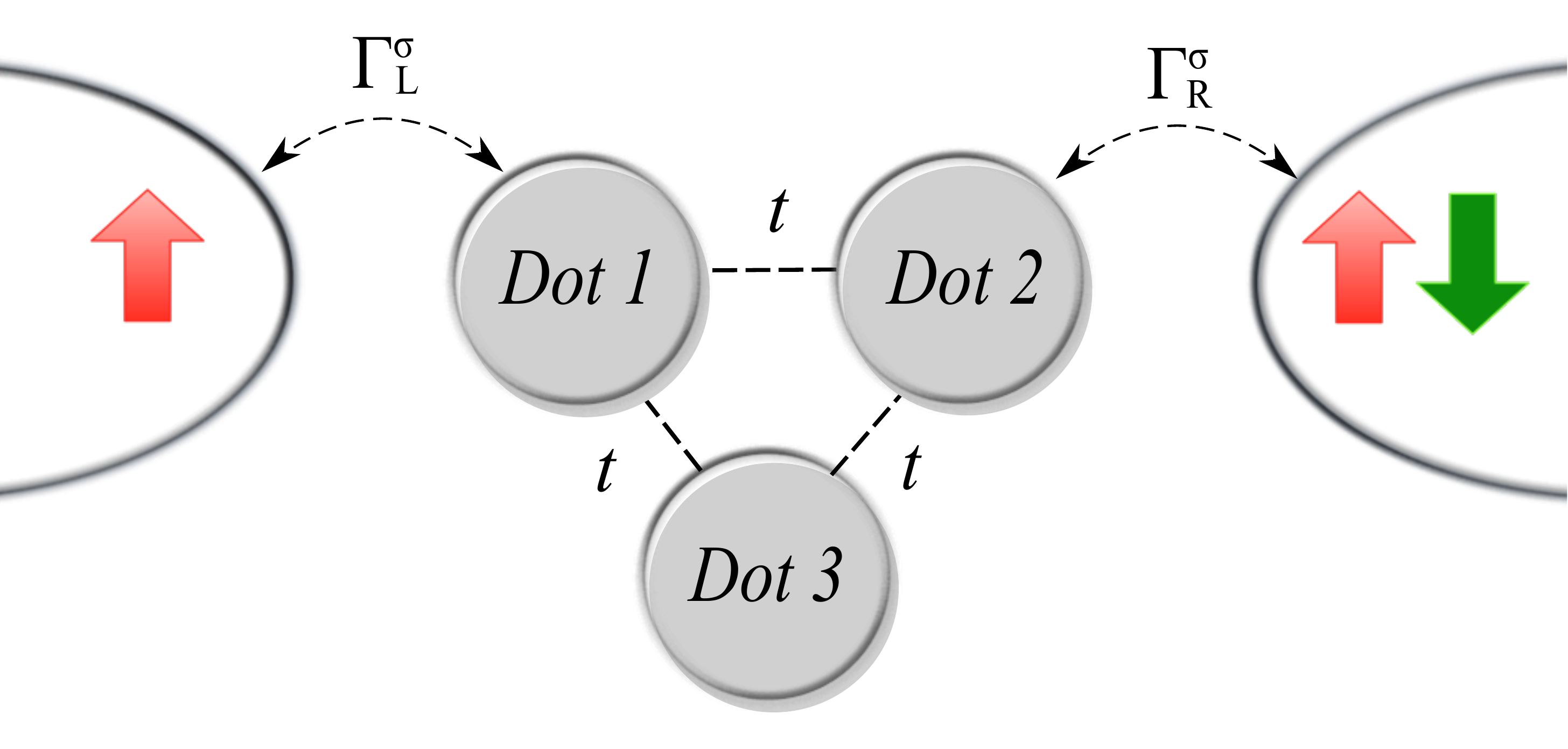}
  \caption{\label{Fig:1}
  Schematic of a triangular quantum dot system coupled to
  ferromagnetic leads. The dot 1 (2) is attached to the left (right) lead with coupling strength $\Gamma^\sigma_\alpha$,
  while the dots are coupled to each other via the hopping matrix elements $t$.}
\end{figure}

The schematic of the considered system
built of three single-level quantum dots forming a triangular geometry is presented in \fig{Fig:1}.
The dots are coupled to each other via the hopping matrix elements $t$.
The first (second) dot is weakly coupled to the left (right) ferromagnetic electrode
with the respective spin-dependent coupling strength $\Gamma^\sigma_\alpha$.
We consider two collinear magnetic configurations of the electrodes:
the parallel (P) one, where both leads' magnetizations point
in the same direction (two red arrows in \fig{Fig:1})
and the antiparallel (AP) one, where the magnetizations point in opposite direction
(red and green arrows). The change of system's magnetic configuration
can be triggered upon applying a weak external magnetic field,
provided the coercive fields of ferromagnets are different.
This field is weak enough, such that the Zeeman energy
associated with this field can be neglected.

The total Hamiltonian of the system is given by
\begin{equation}\label{Eq:Hamiltonian}
    H = H_{\rm Leads} + H_{\rm TQD} + H_{\rm T},
\end{equation}
where the first term
\begin{equation}\label{Eq:HamiltonianLeads}
H_{\rm Leads} =
  \sum_{\alpha=L,R}\sum_{k\sigma}\varepsilon_{\alpha k\sigma} c^\dagger_{\alpha k\sigma} c_{\alpha k\sigma},
\end{equation}
describes the left and right ferromagnetic leads in the noninteracting quasiparticle approximation.
Here, the operator $c^\dagger_{\alpha k\sigma}$ is the creation
operator of an electron with spin $\sigma$,
momentum $k$ and energy $\varepsilon_{\alpha k\sigma}$
in the left or right ($\alpha=L,R$) electrode.
The second term of the Hamiltonian models the triple quantum dot and reads
\begin{eqnarray}\label{Eq:HamiltonianTQD}
  H_{\rm TQD} &=&
  \sum_{j\sigma}\varepsilon_{j} n_{j\sigma}
  + U_j \sum_{j} n_{j\uparrow}n_{j\downarrow}
  +\frac{U_{ij}}{2} \sum_{<ij>}\sum_{\sigma\sigma^\prime}
  n_{\rm i\sigma}n_{\rm j\sigma}
  \nonumber\\
  &+&
  \sum_{<ij>} \frac{t_{ij}}{2} \sum_\sigma(d^\dagger_{i\sigma}d_{j\sigma}+
  d^\dagger_{j\sigma}d_{i\sigma})
 \,.
\end{eqnarray}
The on-site energy is given by $\varepsilon_{j}$, with
$n_{j\sigma}=d^{\dagger}_{j\sigma}d_{j\sigma}$ and
$d^{\dagger}_{j\sigma}$ being the creation operator of an electron
with spin $\sigma$ in the $j$th quantum dot.
The intra- and inter-dot Coulomb interactions are of
strength $U_j$ and $U_{ij}$, respectively.
The hopping between the dots $t$ is assumed to be equal between each pair of the dots.

The last term of the Hamiltonian accounts for the tunneling
between TQD and the leads, and it takes the standard form
\begin{eqnarray}\label{Eq:HamiltonianTunnel}
  H_{\rm T} &=&
  \sum_{k\sigma} (v_L c^\dagger_{Lk\sigma}d_{1\sigma} + v_R c^\dagger_{Rk\sigma}d_{2\sigma}
  + {\rm H.c.})
 \,,
\end{eqnarray}
where $v_L$ and $v_R$ are the tunnel matrix elements between
the left and right leads and the corresponding dots.
The dot-lead coupling strength is given by
$\Gamma^{\sigma}_{\alpha} = 2\pi|v_{\alpha}|^2\rho^\sigma_{\alpha}$, with $\rho^\sigma_{\alpha}$ being
the spin-dependent density of states of lead $\alpha$.
Using the definition of spin polarization of ferromagnetic lead $\alpha$,
$p_{\alpha}=(\rho^+_{\alpha}-\rho^-_{\alpha})/(\rho^+_{\alpha}+\rho^-_{\alpha})$,
the couplings can be written as, $\Gamma^{\pm}_{\alpha}=\Gamma_{\alpha}(1 \pm p_{\alpha})$,
for the spin majority ($\sigma=+$) or minority ($\sigma=-$) subband,
where $\Gamma_{\alpha}= (\Gamma^+_{\alpha} + \Gamma^-_{\alpha})/2$.
We assume equal left and right coupling strengths,
$\Gamma_L = \Gamma_R \equiv \Gamma$.
The applied bias voltage is also assumed to
be symmetrical, $\mu_L=eV/2$ and $\mu_R=-eV/2$.

\subsection{Method}\label{method_sub}

The spin-resolved transport properties of considered system are calculated with the aid of the real-time
diagrammatic technique \cite{diagrams1,diagrams2,thielmann,weymannPRB08}. This method relies on
perturbation expansion of the reduced density matrix and the corresponding operators
with respect to the coupling strength $\Gamma$.
Here we perform all calculations including
the first order, accounting for sequential tunneling processes,
and the second order of expansion, which describes cotunneling.

The reduced density matrix in the steady state can be found from \cite{diagrams1}
\begin{equation}\label{Eq:WMatrix}
    \mathbf{W p} = 0,
\end{equation}
with the normalization condition, Tr$\mathbf{\left\{p\right\}}=1$.
In the above equation $\mathbf{W}$ is a matrix,
the elements $W_{\chi \chi'}$ of which describe transitions between
the TQD many-body states $|\chi \rangle$ and $|\chi' \rangle$,
while $\mathbf{p}$ denotes the probability vector.
The states $|\chi \rangle$ are the eigenstates
of $H_{\rm TQD}$ obtained from the numerical solution of the eigenvalue problem.
Note that the triple dot Hamiltonian, $H_{\rm TQD}$, is not diagonal
in the local occupation basis. However,
in order to explain the microscopic mechanism of the dark states
and the blocking of transport through the system,
we will often express the states $|\chi \rangle$ as superpositions of local occupation states.
Therefore, we assume that a ket in local occupation basis represents occupation of consecutive dots
in the following way: $|\chi_1\chi_2\chi_3 \rangle$, where $\chi_j$
denotes the allowed ($0, \sigma$ and $d$) local states, which stand for empty,
spin $\sigma$ and doubly occupied dot $j$, respectively.
Moreover, to distinguish between different states of the system,
we will additionally use the quantum numbers corresponding
to the total charge $Q$ and spin $z$th component $S_z$ of the TQD, $|Q, S_z \rangle$.

The elements of matrix $\mathbf{W}$ are exactly related to self-energies,
$\Sigma_{\chi \chi'}= iW_{\chi' \chi}$,
which can be determined diagrammatically order by order in $\Gamma$
\cite{diagrams1, diagrams2, thielmann,weymannPRB08}.
A given order in $\Gamma$ corresponds to the respective number of tunneling lines
in diagrams, therefore to find the first and second order contributions,
we consider all topologically different, irreducible diagrams with one and two tunneling lines.
An exemplary calculation of two different diagrams can be found in the Appendix.
The perturbation expansion of the matrix $\mathbf{W}$ starts in the first order in $\Gamma$,
while that of $\mathbf{p}$ starts in the zeroth order.
The corresponding probabilities can be found from the following kinetic equations \cite{diagrams1}
\begin{equation}\label{Eq:PFirstMatrix}
    \mathbf{W}^{(1)}\mathbf{p}^{(0)}=0,
\end{equation}
 and
\begin{equation}\label{Eq:PSecondMatrix}
    \mathbf{W}^{(2)}\mathbf{p}^{(0)}+\mathbf{W}^{(1)}\mathbf{p}^{(1)}=0,
\end{equation}
including ${\rm Tr}\left\{\textbf{p}^{(n)}\right\}=\delta_{0,n}$.

The current flowing through the system can be calculated from \cite{thielmann}
\begin{equation}\label{Eq:Current}
 I = \frac{e}{2\hbar} {\rm Tr} \{ \textbf{W}^I\textbf{p}  \},
\end{equation}
where $\mathbf{W}^I$ is the self-energy matrix, which
accounts for the number of electrons transferred through the TQD system.
For the current we again perform the perturbation
expansion, such that the current in the first order is given by
\begin{equation}\label{Eq:Current1}
 I^{(1)} = \frac{e}{2\hbar} {\rm Tr} \{ \textbf{W}^{I(1)}\textbf{p}^{(0)}  \},
\end{equation}
while the second-order current can be found from
\begin{equation}\label{Eq:Current2}
 I^{(2)} = \frac{e}{2\hbar} {\rm Tr} \{ \textbf{W}^{I(2)}\textbf{p}^{(0)} +  \textbf{W}^{I(1)}\textbf{p}^{(1)}\}.
\end{equation}
The total current, i.e. the first-order (sequential) plus the second-order (cotunneling) current
is then simply given by
\begin{equation}\label{Eq:Current1p2}
 I= I^{(1)}+I^{(2)}.
\end{equation}

In addition to the current we also study the tunnel magnetoresistance,
which describes the change of system's transport properties
when the magnetic configuration of the device is varied.
The TMR can be defined as \cite{julliere, barnas98, Weymann05, barnas08}
\begin{equation}\label{Eq:TMR}
 \mathrm{TMR} = \frac{I^P - I^{AP}}{I^{AP}},
\end{equation}
where $I^P(I^{AP})$ denotes the current flowing through the TQD system in
the parallel (antiparallel) magnetic configuration of ferromagnetic leads.

Finally, we also determine the zero-frequency shot noise $S$
and the corresponding Fano factor ${\rm F} = S/(2|eI|)$,
describing the deviation of the shot noise from the Poissonian value,
$S_P=2|eI|$. A detailed description of how to compute the current fluctuations
within the real-time diagrammatic technique
in a given order of expansion can be found
in Ref. \cite{thielmann}.
By comparing the shot noise to the Poissonian noise $S_P$, one can obtain
an additional information about the statistics of tunneling processes,
which is not contained in the average current \cite{blanterbuttiker}.
In particular, for $F<1$, the shot noise is sub-Poissonian and
its reduction is related to antibunching of tunneling events,
which are correlated by the charging effects.
On the other hand, when $F>1$,
the noise is super-Poissonian and is associated with some bunching mechanism,
e.g. due to the Coulomb blockade \cite{thielmann, blanterbuttiker}.

We note that in order to perform the perturbation expansion,
the coupling strength $\Gamma$ is assumed to be the smallest energy scale in the problem.
Therefore, the approximations made here allow us to study only the weak coupling limit,
while the higher-order correlations,
such as those leading to the Kondo physics \cite{Kondo_Prog.Theor.Phys32/1964,
Goldhaber_Nature391/98,Hewson_book}, are not captured.
Nevertheless, the obtained results are reliable above the exponentially
small Kondo temperature in wide range of finite bias and gate voltage,
which makes this analysis relevant for present and future experimental
investigations of transport through multi-quantum dot systems.


\section{Results and discussion} \label{results}


In this section, we analyze the transport properties of the considered system
for a set of parameters allowing for the formation of one- and two-particle
dark states in the TQD. We want to emphasize that there are several possible
means to obtain dark states in such systems. An important factor is to
distinct one of the relevant quantum dots by detuning its parameters from
the remaining two dots. For instance, this can be obtained by dot-$j$ energy
level detuning $\varepsilon_j=\varepsilon \pm \Delta \varepsilon$,
which was already considered \cite{weymannPRB83}.
Here, we follow the approach proposed by C. P\"oltl \textit{et al}. \cite{emaryPRB80},
where the formation of dark states is conditioned
by an appropriate adjustment of Coulomb interactions,
while the dots' energy levels are the same, $\varepsilon_j=\varepsilon$,
and all the interdot hoppings are also equal.
Experimentally, such setup can be achieved
by appropriate tuning of the dot's size and proper position arrangement.

\begin{figure}[h]
  \includegraphics[width=1\columnwidth]{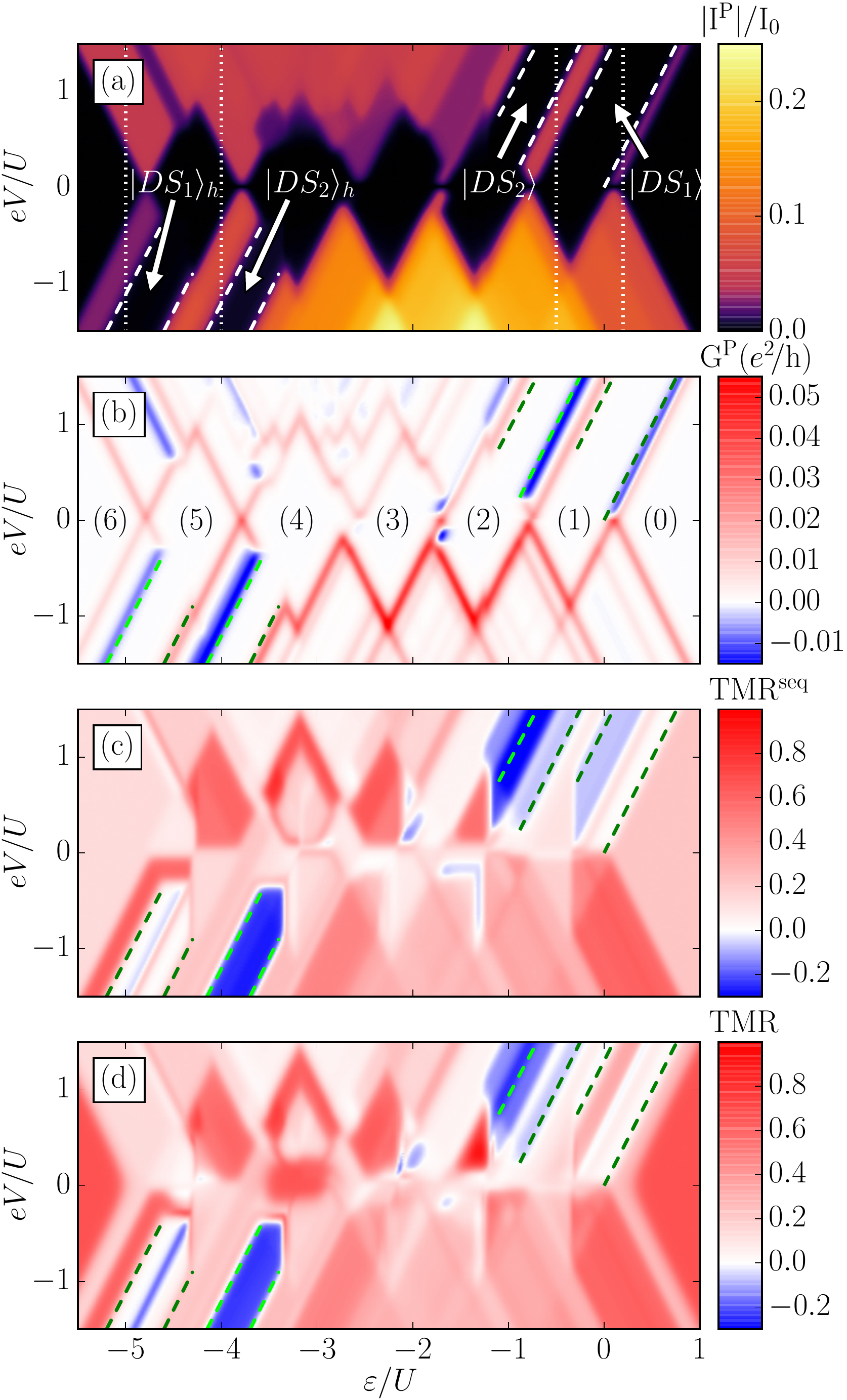}
  \caption{\label{Fig:2}
  (a) The absolute value of the current and (b) the differential conductance in the case of parallel configuration,
  (c) the sequential and (d) total (sequential plus cotunneling) TMR calculated
  as a function of bias voltage and the position of the dots' levels $\varepsilon_j=\varepsilon$.
  The parameters are $U_1=U_2=U_3=U_{13}=U$, $U_{12}=U_{23}=U-\Delta$, where $\Delta = U/5$,
  $t=0.1$, $\Gamma=0.01$, $T=0.03$ in units of $U\equiv 1$, and $p=0.5$.
  The current is plotted in units of $I_0 = e\Gamma/\hbar$.
  The dotted vertical lines in (a) indicate the cross sections
  related with dark states discussed in further sections.
  Dashed lines together with arrows show
  the ranges of voltages, where the current is blocked due to the formation of dark states.
  The numbers in brackets in (b) indicate
  the TQD total occupation number in consecutive Coulomb blockade regimes.}
\end{figure}

The absolute value of the current
in the parallel magnetic configuration with the corresponding differential conductance,
as well as the sequential and total (sequential plus cotunneling) TMR are shown in \fig{Fig:2}
as a function of the bias voltage $V$ and the position of the dots' levels $\varepsilon_j=\varepsilon$.
Since $\varepsilon$ can be tuned experimentally by a gate voltage,
this figure effectively presents the gate and bias voltage dependence of transport characteristics.
Because a typical transport behavior of TQD systems is already relatively well known
\cite{Rogge2008, Granger2010, hsieh12, wrzesniewskiPRB15},
we will mainly point out the differences due to the chosen set of parameters
of discussed model, and especially focus on the effects related to the formation of dark states.

The triple quantum dot, as a multilevel system,
is characterized by a relatively complex Coulomb diamond pattern.
A first general observation is that, due to different Coulomb interactions relevant for the second dot ($j=2$),
which is coupled to the right lead (cf. \fig{Fig:1}),
the magnitude of the current is not the same in both directions with respect
to the applied bias voltage, see \fig{Fig:2}(a).
The current flowing in the positive bias voltage direction is significantly lower, contrary to the current
flowing in the opposite direction.
For low bias, the pattern exposes strong Coulomb blockade regimes
with easily distinguishable number of electrons occupying the TQD,
see \fig{Fig:2}(b).

Assuming that the hopping between the dots is much smaller than
the corresponding inter and intra-dot correlations,
one can estimate the energies at which the occupation of TQD changes
and there is a resonant peak in the linear conductance.
In particular, for $\varepsilon\gtrsim0$ ($\varepsilon \lesssim -5U+2\Delta$), the TQD system is unoccupied
(fully occupied with $6$ electrons).
When $-U+\Delta \lesssim \varepsilon\lesssim 0$,
there is a single electron on the TQD,
for $-2U+\Delta \lesssim \varepsilon\lesssim -U+\Delta$,
there are two electrons on the triple dot
and for $-3U+2\Delta \lesssim \varepsilon\lesssim -2U+\Delta$
the TQD is triply occupied.
On the other hand, when $-4U+2\Delta \lesssim \varepsilon\lesssim -3U+2\Delta$
($-5U+2\Delta \lesssim \varepsilon\lesssim -4U+2\Delta$),
the TQD is occupied by $4$ ($5$) electrons.
The respective electron numbers are indicated in brackets in \fig{Fig:2}(b).

The most interesting features visible both in the current and the differential conductance
are four extended regions of current blockades,
where $I \approx 0$, outgoing from the $1$-, $2$-, $4$- and $5$-electron Coulomb blockade regimes.
The current is suppressed along the energy levels of formed dark states,
as they enter the transport window,
and strong negative differential conductance lines visible in \fig{Fig:2}(b)
are the signatures of that transport phenomena. The presence of dark states also introduces
a strong asymmetry in the bias dependence of the current, resulting in
substantial rectifying properties of the system.

Two of these dark states are formed for TQD energy levels $\varepsilon$
close to the Fermi energy ($\varepsilon = 0$)
and are accessible by applying the bias voltage.
The dots' occupation numbers in those dark states
are respectively equal to $1$ and $2$. For the following discussion,
we therefore label the corresponding states as $1$- and $2$-electron dark states:
$|DS_1\rangle$ and $|DS_2\rangle$.
On the other hand, the two blockades emerging for $\varepsilon\lesssim -4U+2\Delta$,
see \fig{Fig:2}, are related to the formation
of $2$- and $1$-hole dark-states: $|DS_2\rangle_{\!h}$ and $|DS_1\rangle_{\!h}$,
which are symmetric to the aforementioned electron dark states under the particle-hole transformation.
The occupation number of the TQD in these states is equal to $4$ and $5$ electrons, respectively,
however it is more convenient to analyze the opposite-direction hole transport in those transport regimes.
It is also important to notice that the electron dark states
are formed for the opposite sign of the applied voltage bias, as compared to the hole dark states,
see \fig{Fig:2}.
The detailed description of these electron and hole dark states
is the main content of the following subsections.

In \fig{Fig:2} we present the current and differential conductance
only in the case of parallel magnetic configuration. The transport behavior is
qualitatively very similar in the antiparallel configuration
and the differences are well captured by the TMR, which is shown
in \fig{Fig:2}(d). To elucidate the role of cotunneling in transport,
in \fig{Fig:2}(c) we also show the TMR obtained using only the sequential tunneling processes.
First of all, one can note that the TMR behavior within the dark state regimes is quite non-trivial.
While for a wide range of bias and gate voltages,
one observes a typical spin valve behavior with $|I^P|>|I^{AP}|$,
resulting in $\rm{TMR}>0$ \cite{barnas08}, this is not the case
in the dark state regimes.
It can be seen that within the $1$-electron and $1$-hole dark states,
the TMR takes relatively small values in the sequential tunneling approximation
and is strongly modified by cotunneling [cf. Figs. \ref{Fig:2}(c) and (d)],
which can lead to a sign change of the TMR.
On the other hand, both $2$-electron and $2$-hole dark state regions are characterized by
large negative TMR, $\mathrm{TMR}\approx-0.25$,
which is much less affected by second-order processes.
In fact, these two dark states generate the most significant regions of negative TMR in the system,
which implies that, quite counterintuitively, the antiparallel current
is exceeding the current flowing in the parallel configuration, $|I^P| < |I^{AP}|$.
This behavior is explained in detail in the following subsections.
Finally, it is also important to note that cotunneling strongly
modifies the TMR value for the empty (fully occupied) system,
i.e. for low bias and $\varepsilon\gtrsim 0$ ($\varepsilon\lesssim -5U+2\Delta$).
In these transport regions the TMR reaches the Julliere's value \cite{julliere},
which is due to the presence of only elastic cotunneling processes \cite{Weymann05}.
One then finds $\rm{TMR} = \rm{TMR^{Jull}}=2/3$ for considered spin polarization $p=0.5$.

\subsection{One-electron dark state}\label{subs_ds1}

\begin{figure}[t]
  \includegraphics[width=1\columnwidth]{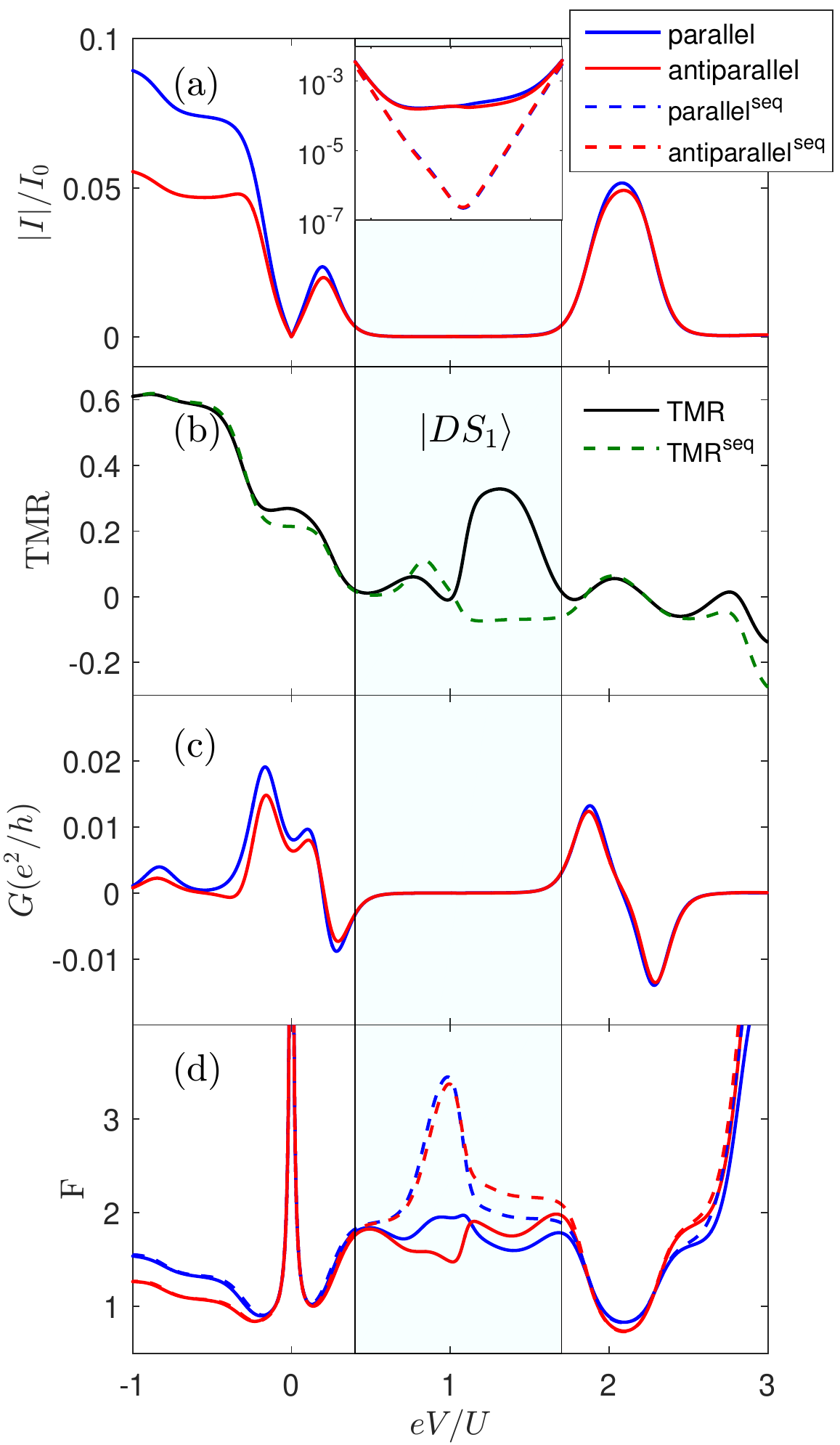}
  \caption{\label{Fig:3}
  The bias voltage dependence of (a) the current,
  (c) the differential conductance and (d) the Fano factor
  in the parallel and antiparallel magnetic configurations
  as well as (b) the TMR in the transport regime
  where $1$-electron dark state $|DS_1 \rangle$ forms.
  The inset in (a) presents the close-up of the system's behavior
  in the dark state regime, where the sequential and total currents are shown on the logarithmic scale.
  The dashed (solid) lines correspond to sequential (sequential plus cotunneling) results.
  The dots' energy levels are $\varepsilon/U=0.2$,
  while the other parameters are the same as in \fig{Fig:2}.
  Note that in order to obtain a $1$-electron dark state
  we introduced a small detuning in the
  hopping integral between the first and third dot,
  $t_{13}=t-\delta t$, with $\delta t = 10^{-2} t$.}
\end{figure}

In this section we analyze in a greater detail the system's transport behavior
for such gate voltages where the $1$-electron dark state $|DS_1\rangle$ is formed.
Before proceeding with discussion
of distinct transport features, let us first make a comment
regarding the choice of parameters.
As already mentioned, to observe dark states it is crucial to
introduce some asymmetry between the dots.
In the present paper this is obtained by allowing for different
Coulomb correlations between the dots.
However, contrary to multi-electron states,
the $1$-electron states are not influenced by Coulomb correlations.
Consequently, the assumed asymmetry in Coulomb interactions cannot result
in breaking of the symmetry of electronic density distribution in the TQD
in the one-electron regime.
Therefore, in order to generate an appropriate dark state within the $1$-electron sector
of the TQD Hamiltonian, we introduced a very small detuning
to the hopping parameter between the first and third dot,
$t_{13}=t-\delta t$, with $\delta t = 10^{-2} t$.
This fine-tuning of parameters suffices to find
a dark state in the $1$-electron parameter space.
We note that in an experimental setup,
it is often of great difficulty to prepare such complex system
in perfect symmetry, consequently, it should be possible
to satisfy the condition favoring the formation of dark states.
We also notice that this small detuning does not affect the other
dark states, which form due to asymmetry introduced by difference
in corresponding Coulomb correlations.

The bias voltage dependence of the current,
differential conductance and the Fano factor in both parallel and antiparallel magnetic configurations
as well as the TMR is shown in \fig{Fig:3}.
In the absence of applied voltage, the system's ground state
is given by $|Q\!=\!0, S_z\!=\!0\rangle=|000\rangle $
and the TQD is empty.
By increasing the positive bias voltage, $eV>0$, the first one-electron
state enters the transport window. It is the following spin doublet,
$|Q\!=\!1, S_z\!=\!\pm \frac{1}{2} \rangle
= \sqrt{\frac{2}{3}}|0 \sigma 0\rangle - \frac{1}{\sqrt{6}}(|\sigma00\rangle + |00\sigma\rangle)$.
This state is a superposition of a single electron delocalized over all the three dots.
It is important to note here that both left and right leads
are coupled to the dots with finite electron occupation.
Therefore, the current can flow through the system and
transport takes place by tunneling processes
between the above-mentioned excited state $|Q\!=\!1, S_z\!=\!\pm \frac{1}{2} \rangle$
and the empty TQD state.
As a result, there is a peak in the current for $eV/U\approx 0.25$,
i.e. for voltages at which the first excited state enters the transport window,
preceded by an associated differential conductance peak, see \fig{Fig:3}(c).

\begin{figure}[t]
  \includegraphics[width=.8\columnwidth]{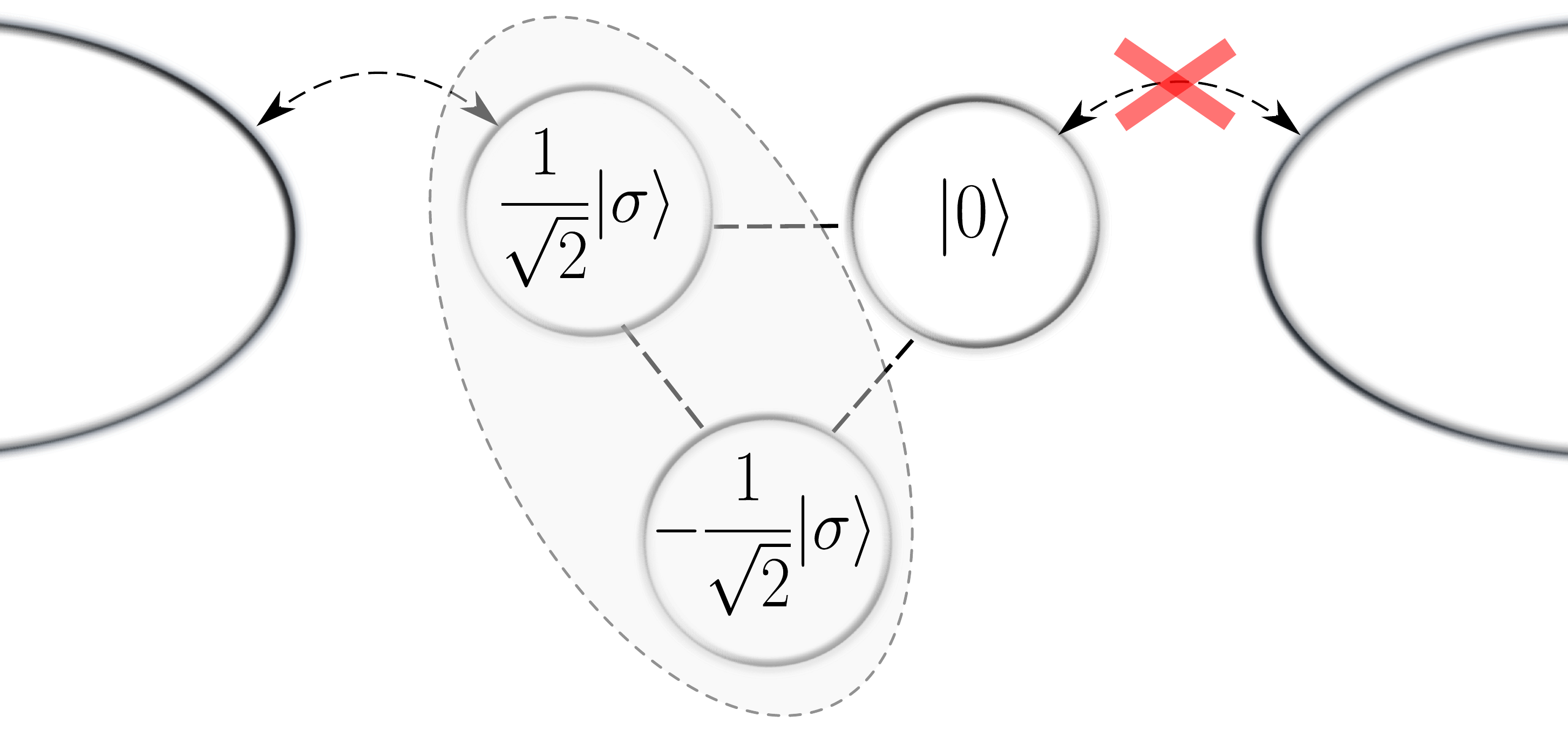}
  \caption{\label{Fig:DS1}
  Graphical representation of the one-electron dark state $|DS_1\rangle$.
  The electron density is distributed between the
  dots $1$ and $3$ leaving the dot $2$ unoccupied and
  thus blocking the transport through the right junction.
}
\end{figure}

Further increase of the applied bias voltage enables the next excited state to enter the transport window,
which is however the following $1$-electron dark state doublet
\begin{equation*}
|DS_1\rangle=|Q\!=\!1,S_z\!=\!\pm \frac{1}{2}\rangle_{DS}=\frac{1}{\sqrt{2}}(|\sigma 0 0\rangle - |0 0 \sigma\rangle).
\end{equation*}
This state is dominating transport in a wide range of bias voltage,
$0.4\lesssim eV/U \lesssim 1.7$, resulting in a strong current suppression
and negative differential conductance right before the current plateau,
see Figs. \ref{Fig:3}(a) and (c).
The form of this dark state is as follows:
the electron occupies evenly the first and third dots, while the amplitude on the second dot is equal to zero.
As the electronic density is completely distributed between the dots $1$ and $3$,
see \fig{Fig:DS1} for a graphical representation of this dark state,
the electron is not able to leave the system through the second
dot and further tunnel to the right electrode coupled with that dot.
It stays trapped in the TQD system completely blocking the current.
Only if the bias voltage is increased above $eV/U \gtrsim 1.7$,
the blockade is lifted as more states enter the transport window.

One can see that the current dependence
shows a similar behavior in both magnetic configurations,
with generally higher absolute values in the parallel configuration
compared to the antiparallel one.
In the dark state region, when only sequential tunneling processes
are included, $\rm{TMR^{seq}}$ is relatively low.
Despite the leads' spin polarization the system stays in
$|DS_1\rangle$ with equal probabilities of both spin components,
and in both configurations the current has a similar value.
When the system is empty, the tunneling of majority spin electron from the left lead is of higher probability.
However, the occupation probabilities of both spin directions
are balanced by the fact, that the tunneling event in opposing direction,
from the TQD back into the left electrode, is also of higher probability
for electron with spin $z$th component aligned with the left electrode spin polarization.
Nevertheless, as can be clearly seen in the inset of
\fig{Fig:3}(a), the sequential processes get exponentially
suppressed in the dark state region, while the dominant contribution
to the current comes from cotunneling, in which electrons
can be transferred between the left and right leads
through virtual states of the system.
Thus, an accurate analysis of the system's spin-resolved transport behavior requires
resorting to the second-order processes.
One can see that cotunneling enhances the TMR in the dark state region,
which has a maximum around $eV/U \approx 1.3$, indicating
that elastic processes are relevant for transport.
In such processes the spin of transferred electron is conserved,
which tends to enhance the magnetoresistive properties of the device.

Let us now discuss the behavior of the Fano factor in the considered
transport regime.
Out of the blockade regime the Fano factor is generally
sub-Poissonian due to the fact the tunneling events are correlated
by Coulomb correlations. In the low bias range,
the Fano factor becomes divergent
since the current vanishes in the zero voltage limit,
while the noise is still finite due to thermal fluctuations.
An interesting behavior can be observed
within the dark state bias window, where
a moderately enhanced super-Poissonian shot noise develops.
When only the first-order processes are included,
the Fano factor reaches values $F\gtrsim 3$ in both magnetic configurations,
see \fig{Fig:3}(d).
However, cotunneling processes reduce this value to $F \approx 2$
in whole blockade range, irrespective of magnetic configuration.
The reduction of the shot noise due to cotunneling
in the dark state region is in agreement with
numerical results obtained earlier for a similar system, but
with nonmagnetic leads \cite{weymannPRB83}.

\subsection{Two-electron dark state}\label{subs_ds2}

\begin{figure}[t]
  \includegraphics[width=1\columnwidth]{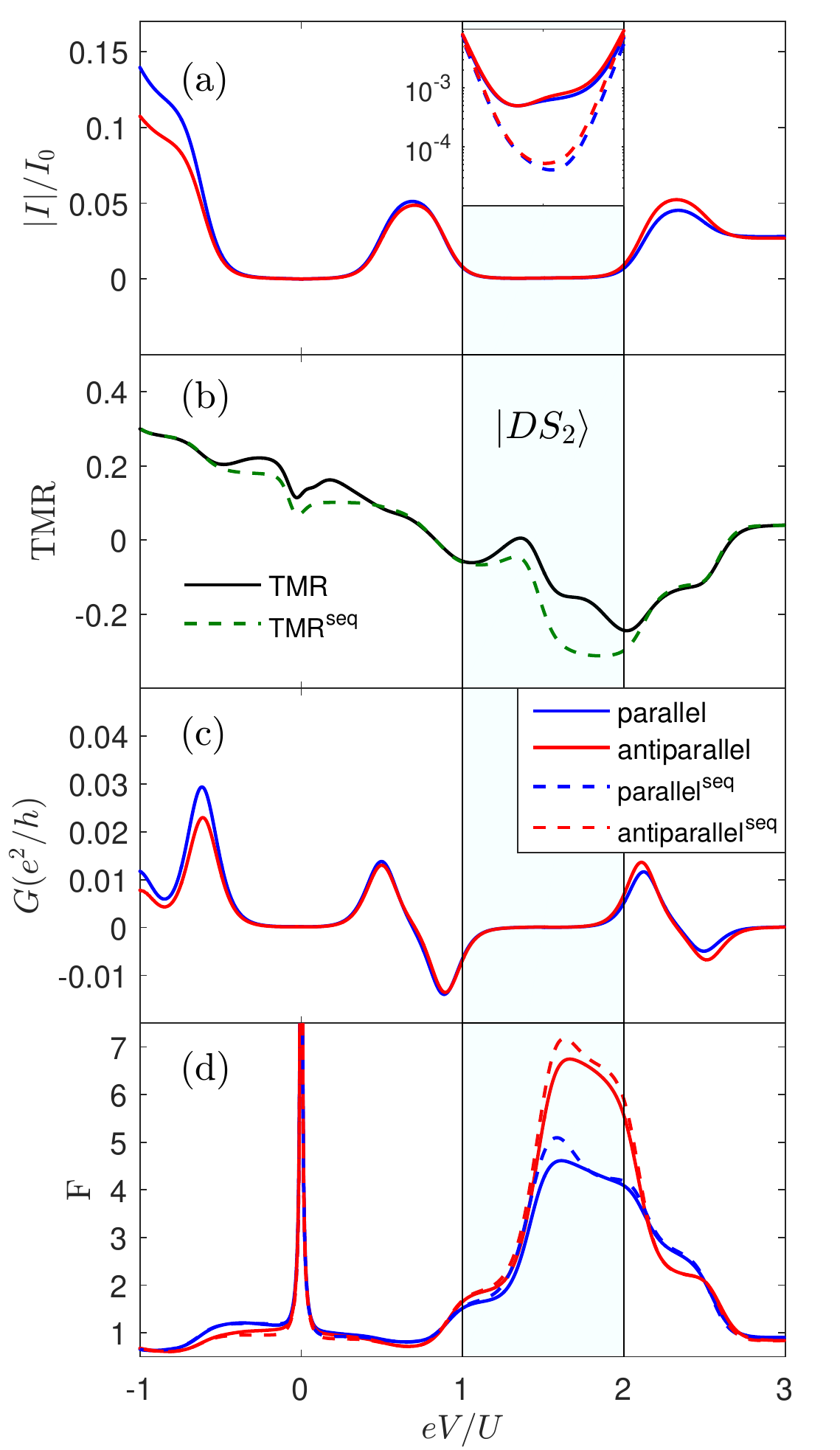}
  \caption{\label{Fig:4}
  The bias voltage dependence of (a) the current,
  (c) the differential conductance and (d) the Fano factor
  in the parallel and antiparallel magnetic configurations
  as well as (b) the TMR in
  in the $2$-electron dark state $|DS_2\rangle$ transport regime.
  The parameters are the same as in \fig{Fig:3} with
  $\varepsilon/U=-0.5$.}
\end{figure}

When the position of the TQD's energy levels is lowered to $\varepsilon/U=-0.5$,
the ground state of the system changes to the one-electron spin doublet state,
${|Q\!=\!1, S_z\!=\!\pm \frac{1}{2} \rangle} = \sqrt{\frac{2}{3}}|0\sigma0\rangle  - \frac{1}{\sqrt{6}}(|\sigma00\rangle + |00\sigma\rangle)$.
By applying the bias voltage in the positive direction,
it is possible to reach a two-electron dark state, which is
responsible for another strong current suppression.
This can be seen in \fig{Fig:4} which presents the
corresponding transport behavior of the system.
First, with increasing the bias voltage, for $eV/U \gtrsim 0.3$,
the current exhibits the first Coulomb step
associated with a two-electron state entering the transport window.
However, around $eV/U \approx 1$, see \fig{Fig:4}(a),
the current suddenly drops due to the following singlet dark state ($|DS_2\rangle = |Q\!=\!2, S_z\!=\!0 \rangle_{DS}$)
\begin{equation*}
|DS_2\rangle \approx \frac{1}{2}(|d 00\rangle + |\!\! \uparrow \!0\! \downarrow\rangle \\
 - |\!\! \downarrow \!0\!  \uparrow\rangle + |0 0 d\rangle)
 \end{equation*}
entering the transport window.
The distribution of electronic density in this state has a similar feature to the $1$-electron dark state,
cf. \fig{Fig:DS1},
i.e. the occupation of the second dot has a vanishing amplitude.
Nonetheless, in the present case, there is a very small but finite amplitude on dot $2$,
allowing for a current leakage, see the inset in \fig{Fig:4}(a).
The presence of dark state again results in a large negative differential
conductance, clearly visible in \fig{Fig:4}(c).

On the other hand, the TMR is a very sensitive quantity
that helps to identify non-trivial behavior within current blockades.
One can see that the TMR behavior in the regime where the system is trapped in the state $|DS_2\rangle$
is different from the $1$-electron dark state ($|DS_1\rangle$) case.
The TMR has a negative sign in almost whole blockade regime,
which also extends slightly above $eV/U \approx 2$,
where the consecutive Coulomb step appears and the new states enter the transport window
allowing for current flow.
The minimum in TMR is well below $\rm{TMR^{seq}}\lesssim-0.3$ in the sequential approximation,
with cotunneling slightly modifying this value to $\rm{TMR}\approx -0.2$,
see \fig{Fig:4}(b)

This particular magnetoresistive behavior can be understood
by taking a closer look at the transport processes occurring
in this regime.
First of all, we note that the negative TMR
is predicted in the sequential approximation
and cotunneling only slightly modifies it.%
\footnote{Note that now the role of cotunneling is not that crucial
as in the case of one-electron dark state, since there is a small leakage
current due to sequential tunneling processes resulting from extremely small but finite occupation
of the second dot.}
Therefore, in order to identify the most important processes for the effect of negative TMR,
let us make a careful analysis of calculated quantities considering sequential tunneling.
The most important factor in this case is how the reduced density matrices for both
magnetic configurations vary between each other.
In the case of parallel configuration, the occupation probability of $|DS_2\rangle$
is close to unity in whole bias range of the blockade,
$p_{|DS_2\rangle} \rightarrow1$. However, this is not the
case for the antiparallel configuration,
where with an increase of the bias voltage,
there is another state,
$|Q\!=\!2, S_z\!=\!\pm 1\rangle =
\sqrt{\frac{7}{8}}|\sigma0\sigma\rangle
+ \frac{1}{4}(|\sigma\sigma0\rangle\!-\!|0\sigma\sigma\rangle)$,
that starts to get small non-zero probability.
It is a two-electron state, polarized in the same direction
as the left electrode, with finite electron density on each quantum dot.
We note that the occupation of the second dot is finite in this state,
which eventually increases the current in the antiparallel configuration
compared to the parallel one.

To find the reason for difference in probability distributions,
we need to consider and compare the dominating processes in both magnetic configurations.
In the parallel configuration, when the system
leaves the singlet dark state $|DS_2\rangle$
by removing one of the majority-spin electron,
which is a process of finite, but very small probability,
the TQD remains occupied by a minority-spin electron.
The immediate consecutive tunneling event
brings another majority-spin electron onto
the TQD restoring $|DS_2\rangle$ dark state,
such that the system remains trapped in this state for a relatively long time
and, as a result, $p_{|DS_2\rangle}\rightarrow1$.
In the case of antiparallel configuration,
a more complex tunneling sequence defines the dominating scenario.
Now, the system leaves the two-electron dark state
$|DS_2\rangle$ by tunneling of electron
with spin aligned along the polarization of the right lead,
however, in the antiparallel configuration, this is the opposite spin direction to the magnetization of the left lead.
This event leaves the TQD with the electron of spin aligned
along the magnetization of the left electrode.
Consecutive tunneling of another majority-spin
electron from the left lead is now preferred,
which results in a transition to the state $|Q\!=\!2, S_z\!=\!\pm 1\rangle$,
instead of a transition to $|DS_2\rangle$,
which was the case in the parallel configuration.
In consequence, the probability of the state
$|Q\!=\!2, S_z\!=\!\pm 1\rangle$ is enhanced,
which results in a larger current in the antiparallel configuration
compared to the parallel one, see \fig{Fig:4}(a).

The difference in sequences of the most probable transport processes in both magnetic configurations
is also visible in the behavior of the shot noise, which is shown in \fig{Fig:4}(d).
One can see that now the shot noise is enhanced in
the dark state region compared to the one-electron dark state case.
Moreover, there is a large difference in the Fano factor
in both magnetic alignments.
The Fano factor reaches $F \approx 7$
in the antiparallel configuration and $F \approx 4$
in the case of parallel configuration.
Note also that the influence of cotunneling on the shot noise
is now much smaller compared to the case
shown in \fig{Fig:3}(d), which is due to the reasons discussed above.

\subsection{Two-hole dark state}\label{subs_ds3}

The transport region with values of TQD's energy levels
$\varepsilon/U \lesssim -4$ also displays non-trivial transport characteristics,
see \fig{Fig:2}. The triple dot is then occupied with relatively high number of electrons.
There are two strong current blockades, located
on the opposite sign of applied bias voltage
(compared to previously discussed cases),
which, similarly, are formed due to the presence of the dark states.
Moreover, the dark states in those regimes consist
of four and five electrons trapped in the system.

\begin{figure}[t]
  \includegraphics[width=1\columnwidth]{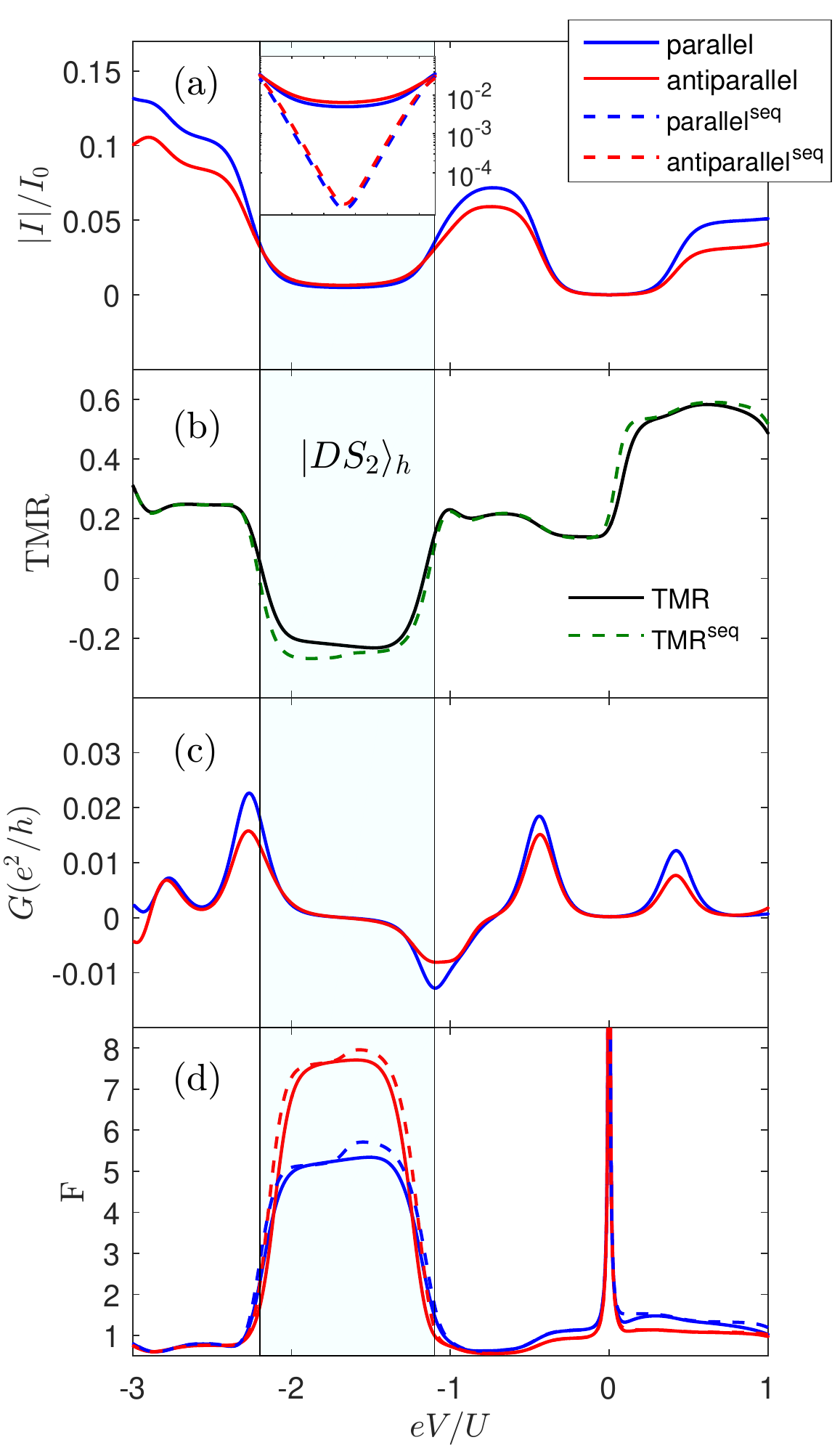}
  \caption{\label{Fig:5}
  The bias voltage dependence of (a) the current,
  (c) the differential conductance and (d) the Fano factor
  in the parallel and antiparallel magnetic configurations
  as well as (b) the TMR in the transport regime
  where the $2$-hole dark state $|DS_2\rangle_{\!h}$ occurs.
  The parameters are the same as in \fig{Fig:3} with $\varepsilon/U=-4$.}
\end{figure}

Let us first discuss the transport regime where
the TQD is occupied by five electrons in the absence of
applied bias. The associated transport characteristics
calculated for $\varepsilon/U = -4$ are presented in \fig{Fig:5}.
The ground state of the system is then given by the following (unnormalized) doublet
state
$|Q\!=\!5, S_z\!=\!\pm \frac{1}{2}\rangle=2t(|\sigma d d\rangle + |d d \sigma\rangle) +
(\sqrt{4\Delta^2+4t\Delta+9t^2}-2\Delta-t)|d \sigma d\rangle$.
The first excited state, which is responsible for the first Coulomb step
in the direction of negative bias voltage is
${|Q\!\!=\!\!4, S_z\!\!=\!\!0\rangle}\!\!=
\!\!\frac{1}{\sqrt{6}}({|0dd\rangle} - {|\!\!\uparrow \! d\!\downarrow\rangle} + {|\!\!\downarrow \!d\!\uparrow\rangle} + {|d d 0\rangle})
+ \frac{1}{\sqrt{12}}({|d\! \uparrow \downarrow\! \rangle} + {|\!\!\downarrow \uparrow \! d\rangle} - {|\!\!\uparrow \downarrow \!d \rangle}
 - {|d \!\downarrow \uparrow \rangle})$.
It is a four-electron state build of eight-component
linear combination of local occupation basis states,
which allows for transport together with the five-electron ground state.
The crucial factor for the charge transport
to happen is that the state ${|Q\!=\!4, S_z\!=\!0\rangle}$ allows for
transitions to the five-electron state
by means of tunneling process of electron from the right lead onto the TQD.
Such processes can happen, when the second dot coupled to the
right electrode is not fully occupied in the considered state.
This is however not the case for the next state,
which enters the transport window with further increase of the bias voltage.
This state results in the current blockade, which appears for negative bias voltage in the range of
$-2.2\lesssim eV/U \lesssim -1.2$, see \fig{Fig:5}(a).
The explicit form of this four-electron singlet dark state
($|DS_2\rangle_{\!h}\!\!=\!\!|Q\!\!=\!\!4,S_z\!\!=\!0\rangle_{DS}$)
is the following
\begin{equation}
|DS_2\rangle_{\!h} \approx \frac{1}{2}(|\!\!\downarrow \!d\! \uparrow\rangle\!-\!|\!\!\downarrow \!d\!\uparrow\rangle
- |dd0\rangle + |0dd \rangle) .
\end{equation}
In this case, when the system is trapped in state $|DS_2\rangle_{\!h}$,
the possibility of transition to the five-electron state is blocked.
Each of the components building this state
contains a fully occupied second dot,
therefore tunneling of electron from the lead
through the right junction is prohibited.
In order to leave this state, the electron has to either tunnel from
the TQD through the left junction,
however the three-electron states are above the transport window,
or tunnel back through the right lead,
which is the event of a very low probability.
Consequently, the system becomes trapped in
the two-hole dark state $|DS_2\rangle_{\!h}$
and the current blockade develops.

It is convenient and more intuitive to analyze the transport properties in this regime
when the TQD states are considered in the hole basis $(h\! -\! basis)$.
With the following electron-hole transformation of local dot's states:
$|0\rangle \rightarrow |d\rangle_{\!h}$ , $|\sigma\rangle \rightarrow |\bar{\sigma}\rangle_{\!h}$
and $|d\rangle \rightarrow |0\rangle_{\!h}$, we can rewrite $|DS_2\rangle_{\!h}$ as
\begin{equation}
|DS_{2}\rangle_{\!h} \approx \frac{1}{2}(|\!\!\uparrow\! 0\! \downarrow\rangle_{\!h} - |\!\!\downarrow \!0\! \uparrow\rangle_{\!h} -
|0 0 d\rangle_{\!h} + |d 0 0 \rangle_{\!h}).
\end{equation}
Now, the blockade can be understood as the effect
of negative interference forming a two-hole dark state,
where the second dot is completely unoccupied
by holes (doubly occupied by electrons).
Then, the second dot is effectively decoupled from
the right lead and the hole transport through this junction is suppressed.

Because the structure of the two-hole dark state
is quite similar to the two-electron dark state,
the transport behavior is qualitatively similar in the two cases,
cf. Figs. \ref{Fig:4} and \ref{Fig:5}.
First of all, the range of the bias voltage
where the dark state dominates transport is of comparable size.
Moreover, the behavior of the TMR is also qualitatively similar,
i.e. in the whole range of bias where the current is suppressed
the TMR has negative values, $\rm{TMR} \approx - 0.2$, see \fig{Fig:5}(b).
The mechanism leading to such behavior is the same as
in the case analyzed in Sec. \ref{subs_ds2}.
The enhanced Fano factor in the dark state regime
reaches $F\approx 8$ in the antiparallel configuration
and is reduced to $F\approx 5$ in the case of parallel configuration.
A subtle difference when comparing the Fano factor behavior
in the case of dark states $|DS_2\rangle$ and $|DS_2\rangle_{\!h}$
is that for the latter case the strongly enhanced
value is present in the whole range of the current blockade.
This is contrary to the two-electron dark state case,
where there is a small range of bias voltage in the blockade
with significantly lower values of the Fano factor, $F\approx 2$,
coincidental with $\rm{TMR}\approx 0$.

\subsection{One-hole dark state}

\begin{figure}[t]
  \includegraphics[width=1\columnwidth]{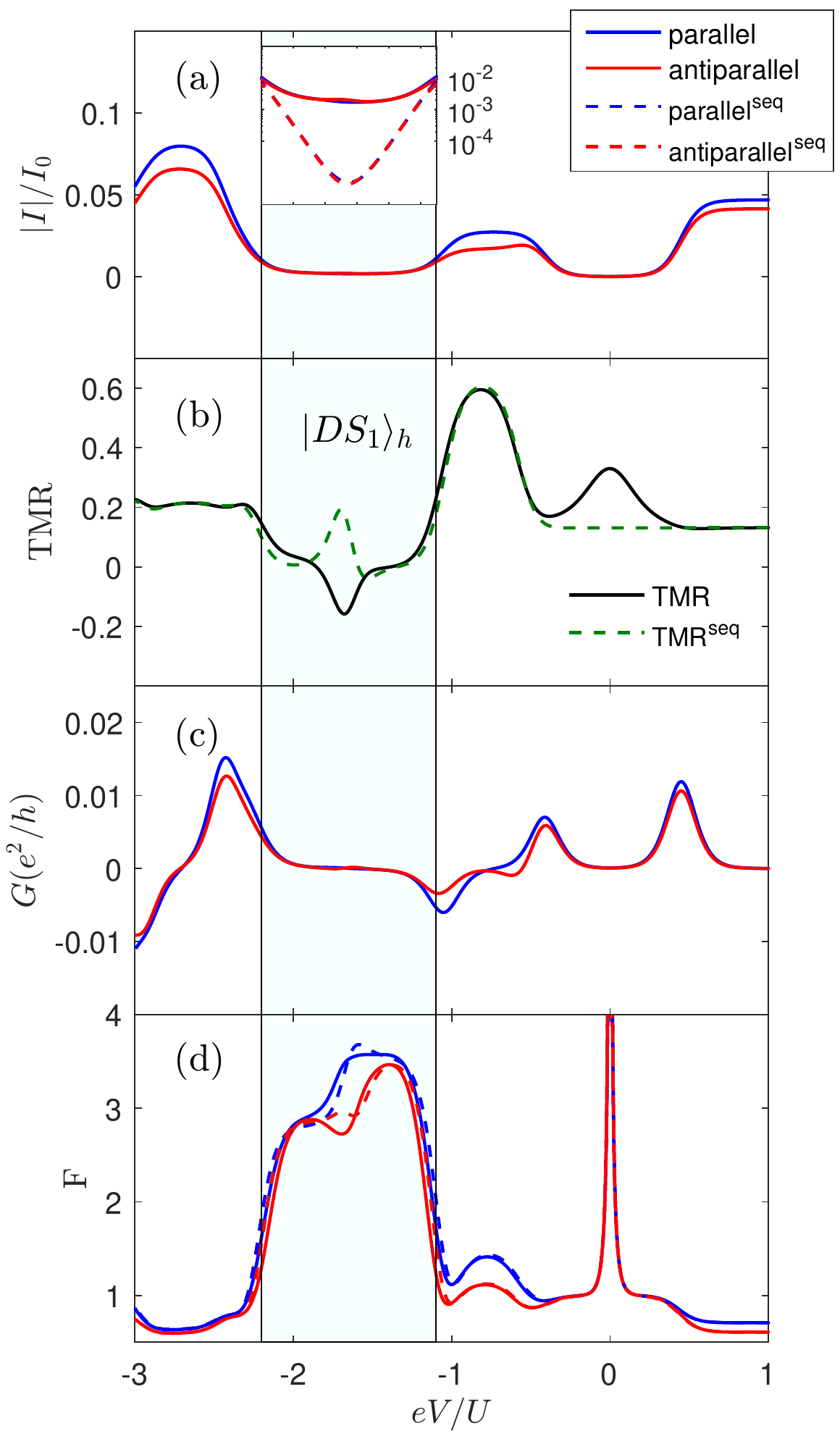}
  \caption{\label{Fig:6}
  The bias voltage dependence of (a) the current,
  (c) the differential conductance and (d) the Fano factor
  in the parallel and antiparallel magnetic configurations
  as well as (b) the TMR in
  in the $1$-hole dark state $|DS_1\rangle_{\!h}$ transport regime.
  The parameters are the same as in \fig{Fig:3} with $\varepsilon/U=-5$.}
\end{figure}

Finally, in this section we discuss the transport regime where
one-hole dark states are formed.
To realize such situation,
we set the TQD energy levels to $\varepsilon/U=-5$,
so that for zero bias the system is fully occupied
and the ground state is
$|Q\!=\!6, S_z\!=\!0\rangle=|d d d\rangle$.
The corresponding transport characteristics are displayed in \fig{Fig:6}.
When applying the negative bias voltage,
the system starts to conduct the current around $eV/U \approx -0.4$
and then, for larger voltages, there is a wide blockade for
$-2.2 \lesssim eV/U \lesssim -1.2$, see \fig{Fig:6}(a).
The dark state responsible for this current suppression is
the following doublet state
$|DS_1\rangle_{\!h}={|Q\!=\!5, S_z\!=\!\pm \frac{1}{2}\rangle_{DS}}
=\frac{1}{\sqrt{2}}(|\sigma d d \rangle - |d d \sigma \rangle)$.
Similarly to the previous case of two-hole dark state, we again see that the
second dot is fully occupied by two electrons.
This configuration blocks the electron flow through the right junction
onto the TQD, which is the promoted direction by the applied bias.
The tunneling processes through the left junction
are now also energetically very unfavorable and, as a result,
the system remains trapped in the dark state $|DS_1\rangle_{\!h}$
blocking the current.
The discussed dark state can be also conveniently written in the hole-basis as
\begin{equation}
|DS_{1}\rangle_{\!h}=\frac{1}{\sqrt{2}}(|\sigma 0 0 \rangle_{\!h} - |0 0 \sigma \rangle_{\!h}).
\end{equation}
It can be now clearly seen that the above one-hole dark state
has a similar form to $|DS_1\rangle$ discussed in Sec. \ref{subs_ds1}.
see also \fig{Fig:DS1}.
Consequently, we find that the blocking mechanism and interference effects are analogous
in the two cases, with the difference that the holes are considered instead of the electrons.

The TMR behavior in the current blockade regime also has some
similarities to the case of $|DS_1\rangle$ dark state.
In both cases the current obtained within the sequential tunneling approximation is relatively low
and the transport behavior is predominantly determined by cotunneling processes.
Furthermore, in the case of one-hole dark state
cotunneling results in a sign change of the TMR
in the middle of the blockade, where the first-order processes resulted in
a maximum, see \fig{Fig:6}(b).
This clearly confirms that the second-order processes are important
and dominate in this transport regime.
Moreover, these processes strongly enhance the current in the antiparallel configuration.
However, the region of negative TMR is now significantly smaller than those predicted
in the case of current blockades caused by the formation
of two-particle dark states $|DS_2\rangle$ and $|DS_2\rangle_{\!h}$.
Finally, one can also see that the Fano factor is consistently
enhanced in the $|DS_1\rangle_{\!h}$ dark state regime, reaching
$F \approx 3.5$ in both magnetic configurations,
see \fig{Fig:6}(d).

\section{Conclusions} \label{conclusions}

In this paper we have studied the influence of dark
states on the spin-resolved transport properties of a
triple quantum dot molecule attached to ferromagnetic contacts.
The considerations were performed by
using the real-time diagrammatic method and considering both sequential
and cotunneling processes.
By optimizing the system parameters,
we have showed that the current flowing through the device
can be blocked due to the coherent population
trapping in a dark state.
We have analyzed the transport behavior in the case
of various dark states in the system,
including one and two-particle (either electron or hole) dark states.
In all those cases we have shown that transport is mainly
determined by cotunneling processes,
which result in a great modification of the magnetoresistive
properties of the system.
In particular, we have demonstrated that the interplay of spin-polarized transport
with two-particle dark states can lead to negative tunnel magnetoresistance.
Moreover, we have found super-Poissonian shot noise
in the current blockade regimes, which can be additionally
enhanced by spin-dependence of tunneling processes.
Finally, we have also indicated that the dark states with high number of electrons
can be conveniently understood and analyzed as states
formed by interference of holes, and the resulting transport characteristics
can be discussed within the hole current framework.
In this respect, we have also emphasized some similarities
between the transport regions with the
electron and hole dark states containing the same numbers of particles.


\begin{acknowledgements}
This work was supported by the Polish National Science
Centre from funds awarded through the decision No. DEC-2013/10/E/ST3/00213.
\end{acknowledgements}


\section*{Appendix: Details of calculations} \label{appendix}

In this appendix we present the details
of the performed calculations with the aid of the real-time diagrammatic technique approach.
In order to solve the kinetic equation (\ref{Eq:WMatrix})
and obtain the density matrix elements,
one has to find the self-energies $\Sigma_{\chi \chi'}$,
which are related to the elements of matrix $\mathbf{W}$ through:
$\Sigma_{\chi \chi'}= iW_{\chi' \chi}$.
The most difficult part of the calculations is to evaluate all irreducible,
topologically different diagrams describing tunneling processes,
which can be done with the aid of the diagrammatic rules
\cite{diagrams1, diagrams2}.
Because in this paper we studied the effects of sequential tunneling and cotunneling,
one needs to determine the self-energies up to the second order
of perturbation expansion in the tunnel coupling $\Gamma$.
In practice, it is necessary to consider all the diagrams
containing one and two tunneling lines.
Below, we present an exemplary contributions from the first and second order diagrams,
as well as their contributions to a one specific self-energy.

\subsection{First-order diagrams}

The first-order diagrams involve a single tunneling line.
Below, we show a diagram contributing to the
following first-order self-energy $\Sigma^{(1)}_{\chi(N) \chi'(N+1)}$.
\begin{eqnarray*}
\includegraphics[width=.6\columnwidth,left]{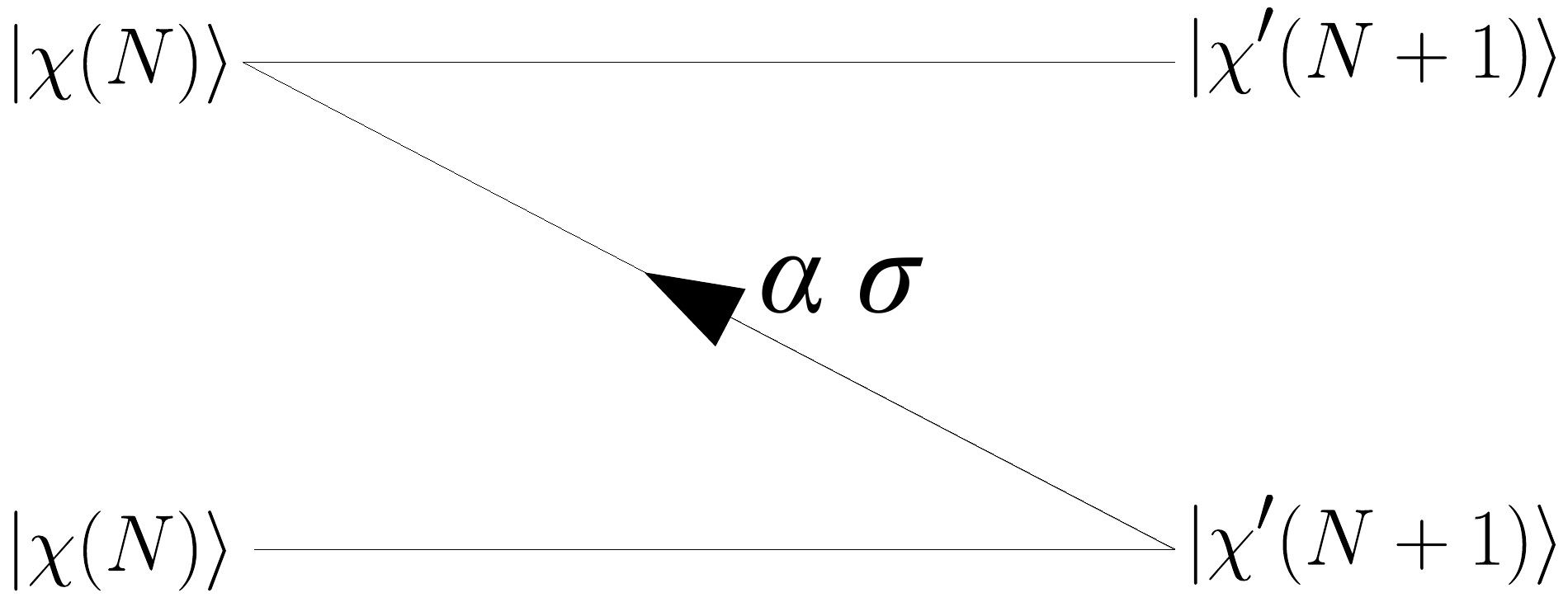}\nonumber\\
= (-1)^1 \sum_{j}
\int \! d \omega \frac{\gamma_{\alpha \sigma}(\omega)}{\omega - \varepsilon_{\chi'} + \varepsilon_{\chi}+i\eta}
| \langle \chi' | d^{\dagger}_{j \sigma} | \chi \rangle |^2,
\end{eqnarray*}
where $\gamma_{\alpha \sigma}=\frac{\Gamma^{\sigma}_{\alpha}}{2\pi}f_{\alpha}(\omega)$
is a factor associated with each tunneling line,
$f_{\alpha}(\omega)$ is the Fermi-Dirac distribution of lead $\alpha$
and $\eta = 0^{+}$.
This diagram corresponds to an electron with spin $\sigma$
tunneling from the lead $\alpha$, between $|\chi(N)\rangle$ and $|\chi'(N+1)\rangle$ states,
where $N$ indicates total occupation number $N=\sum_{j\sigma}n_{j\sigma}$.

When all the topologically different first-order diagrams are evaluated,
the respective self-energies can be determined.
In particular, the self-energy $\Sigma^{(1)}_{\chi(N) \chi'(N+1)}$ is given by
\begin{equation*}
\Sigma^{(1)}_{\chi(N) \chi'(N+1)} = 2\pi i \sum_{\alpha \sigma}\sum_j
\gamma_{\alpha \sigma}(\varepsilon_{\chi'} - \varepsilon_\chi)
| \langle \chi' | d^{\dagger}_{j \sigma} | \chi \rangle |^2.
\end{equation*}

\subsection{Second-order diagrams}

The second-order diagrams involve two tunneling lines.
Here, as an example we present a contribution from a diagram that
contributes to the second-order
self-energy $\Sigma^{(2)}_{\chi(N) \chi'(N+2)}$ due to
two tunneling events of electrons with spins $\sigma$, $\sigma'$,
tunneling from the leads $\alpha$, $\alpha'$. It is given by
\begin{eqnarray*}
\includegraphics[width=.6\columnwidth,left]{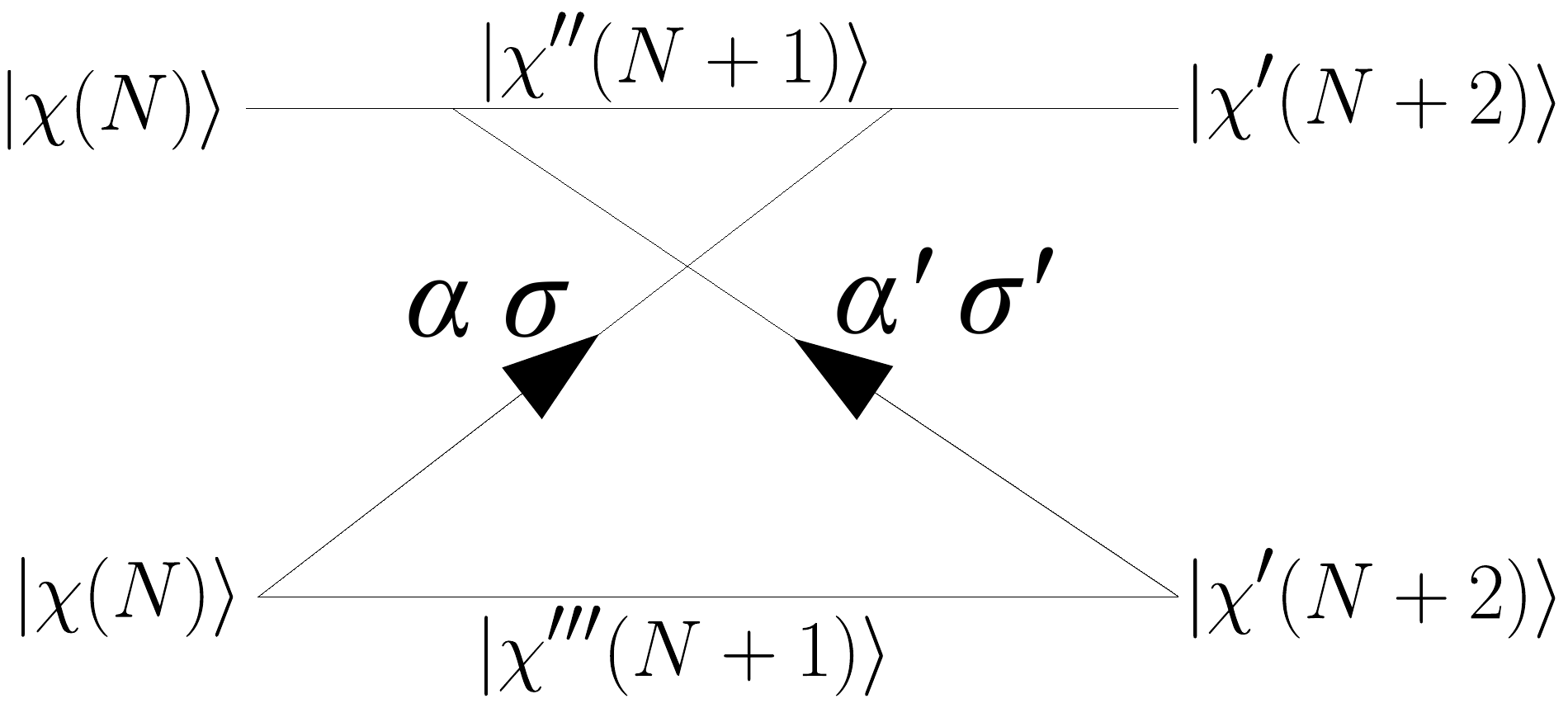}\nonumber\\
\begin{aligned}
\phantom{z} = &(-1)^3 \sum_{\chi'' \chi'''} \sum_{jj'} \int\!\!\!\!\int d \omega_1 d \omega_2
\frac{\gamma_{\alpha \sigma}(\omega_1)}{-\omega_1 - \varepsilon_\chi + \varepsilon_{\chi'''}+i\eta} \\
& \times \frac{\gamma_{\alpha' \sigma'}(\omega_2)}{-\omega_1 + \omega_2 - \varepsilon_{\chi''} + \varepsilon_{\chi'''}+i\eta}
\frac{1}{\omega_2 - \varepsilon_{\chi'} + \varepsilon_{\chi'''}+i\eta} \\
& \times \langle \chi | d_{j \sigma} | \chi''' \rangle \langle \chi''' | d_{j' \sigma'} | \chi' \rangle
\langle \chi' | d^{\dagger}_{j \sigma} | \chi'' \rangle \langle \chi'' | d^{\dagger}_{j' \sigma'} | \chi \rangle,
\end{aligned}
\end{eqnarray*}
An important step simplifying the calculations is to use the mirror rule.
By reflecting any diagram horizontally and changing directions of all tunneling lines, one obtains
the contribution which is an opposite sign complex conjugate of the initial diagram.
The pairs of such symmetric diagrams contribute
only with a summed imaginary parts, while the real parts cancel out.
The integrations in the above formula can be performed analytically by
using the Cauchy's principal value theorem and realizing that
integrals of the form
\begin{equation*}
F_{\alpha \sigma}^\gamma = \int^{\infty}_{-\infty} d\omega \frac{\gamma_{\alpha \sigma}(\omega)}{(\omega - \varepsilon + i\eta)^\gamma}
\end{equation*}
can be evaluated using the digamma function and its derivatives.

Finally, all contributions to the second-order self-energy $\Sigma^{(2)}_{\chi(N) \chi'(N+2)}$
can be graphically represented as a sum of the following diagrams
\begin{equation*}
  \includegraphics[width=1\columnwidth]{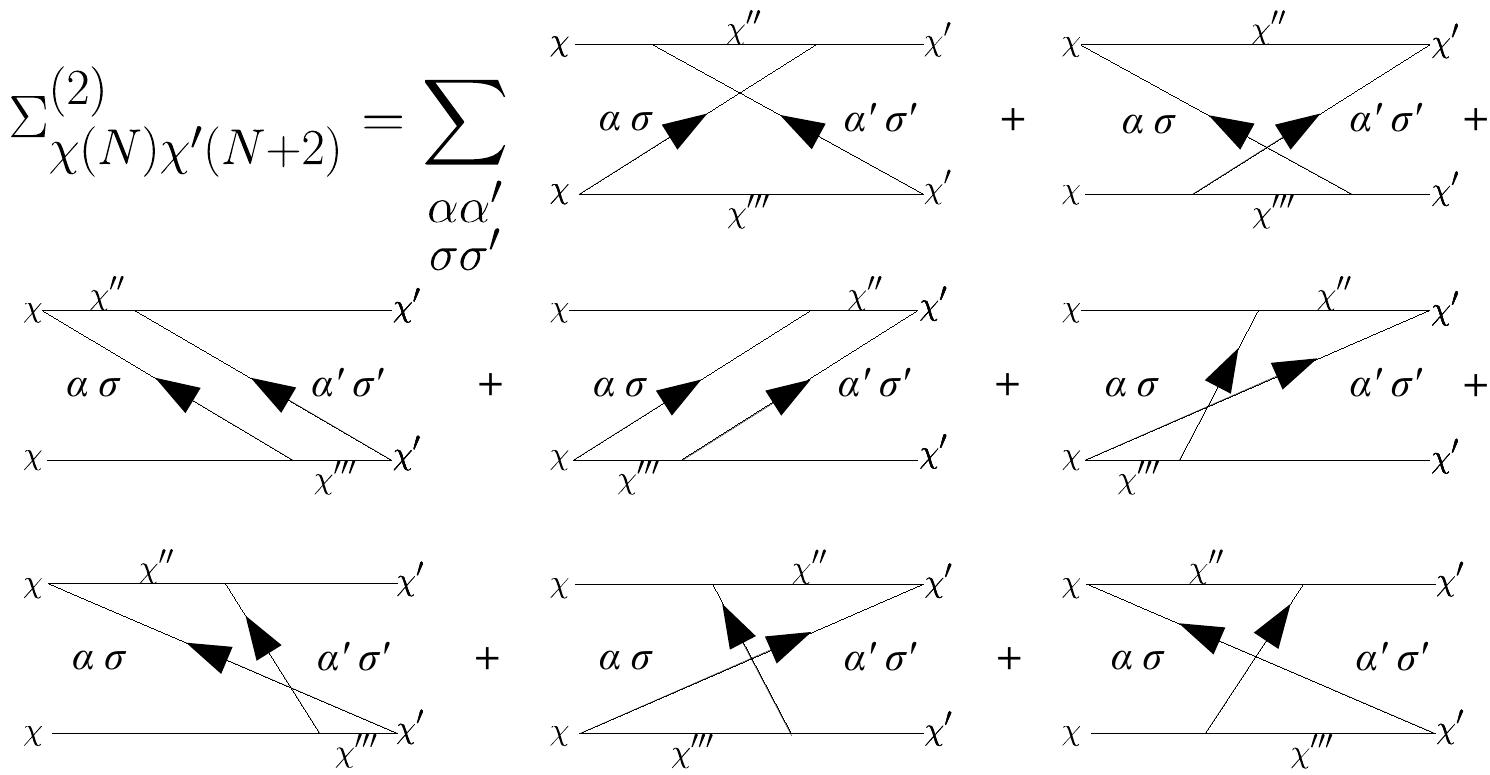}
\end{equation*}


%

\end{document}